# Electric-field control of magnetic domain wall motion and local magnetization reversal

Tuomas H. E. Lahtinen, Kévin J. A. Franke and Sebastiaan van Dijken*

*NanoSpin, Department of Applied Physics, Aalto University School of Science, P.O. Box 15100, FI-00076 Aalto, Finland. *e-mail: sebastiaan.van.dijken@aalto.fi*

**Spintronic devices currently rely on magnetic switching or controlled motion of domain walls by an external magnetic field or spin-polarized current. Achieving the same degree of magnetic controllability using an electric field has potential advantages including enhanced functionality and low power consumption. Here, we report on an approach to electrically control local magnetic properties, including the writing and erasure of regular ferromagnetic domain patterns and the motion of magnetic domain walls, in multiferroic CoFe-BaTiO$_3$ heterostructures. Our method is based on recurrent strain transfer from ferroelastic domains in ferroelectric media to continuous magnetostrictive films with negligible magnetocrystalline anisotropy. Optical polarization microscopy of both ferromagnetic and ferroelectric domain structures reveals that domain correlations and strong inter-ferroic domain wall pinning persist in an applied electric field. This leads to an unprecedented electric controllability over the ferromagnetic microstructure, an accomplishment that produces giant magnetoelectric coupling effects and opens the way to multiferroic spintronic devices.**

In the field of spintronics, researchers have controlled magnetic switching and domain wall motion using spin-polarized currents and these phenomena have formed the basis for magnetic random access memory (MRAM) [1] and magnetic nanowire device concepts [2,3]. Employing significant amounts of electrical current, however, is inevitably accompanied by energy dissipation, and in this context, electric-field induced magnetization reversal without major current flow would be desirable. Multiferroic materials are a promising candidate for the realization of electric-field controlled spintronic devices [4,5], but practical requirements such as strong magnetoelectric coupling and room temperature operation pose significant scientific challenges for single-phase multiferroic materials. Cross-linking between ferromagnetic and ferroelectric order parameters via interface effects in multiferroic heterostructures provides a viable alternative, as it allows for an independent optimization of both ferroic phases [6]. Up to now, several mechanisms for magnetoelectric coupling at ferromagnetic-ferroelectric interfaces have been explored including charge modulation [7-10], exchange interactions [11-17], and strain transfer [18-25]. This has led to great advances in electric-field control of macroscopic magnetic properties (anisotropy, exchange bias, and ferromagnetic resonance frequency) and the demonstration of interface-induced multiferroicity [10]. An extension of this degree of controllability to micromagnetic elements such as magnetic domains and domain walls would constitute another important step towards the realization of practical multiferroic devices, as it would enable electric-field actuation of magnetic functions that up to now have only be addressed by an external magnetic field or spin-polarized current. Previous studies on local control of ferromagnetism using imaging techniques that are sensitive to both ferromagnetic and ferroelectric order parameters have revealed domain correlations and electric-field induced magnetic switching of irregular domains in single-phase multiferroics [26,27] and exchange-biased multiferroic heterostructures [13].

Here, we demonstrate that it is possible to precisely write and erase regular ferromagnetic domain patterns and to control the motion of magnetic domain walls in small electric fields. Our approach consists of two steps. First, a well-defined ferroelastic domain structure is fully transferred from a tetragonal ferroelectric to a magnetostrictive medium via interface strain transfer during thin film growth. Hereafter, an electrical bias voltage is used to reversibly manipulate the original magnetic microstructure by polarization reversal and domain wall motion in the ferroelectric. The results show that microscopic features such as ferroelastic domain history and ferromagnetic-ferroelectric domain wall pinning critically determine the magnitude of macroscopic magnetoelectric coupling effects.

For this study, thin CoFe films were directly grown onto BaTiO$_3$ substrates with a regular ferroelastic stripe pattern that is typified by 90° in-plane rotations of the ferroelectric polarization and elongated $c$ axis (Fig. 1a). This ferroelastic $a_1 - a_2$ domain structure produces a lateral strain modulation of 1.1% ($a = b = 3.992$ Å, $c = 4.036$ Å). CoFe with an atomic percentage of 60% Co and 40% Fe was selected because of its large magnetostriction and small magnetocrystalline anisotropy [28]. This materials choice ensured that local magnetoelastic energies compared favorably to other energy scales in the ferromagnet (exchange, magnetostatic, and magnetocrystalline energies), which is a prerequisite for electric-field control of ferromagnetic microstructures. Both ferromagnetic and ferroelectric domain patterns were imaged using optical polarization microscopy techniques. Real-time measurements of domain correlations in an applied electric field were facilitated by the use of semi-transparent CoFe films with a thickness of 15 nm. More details on sample fabrication and characterization procedures are presented in the Methods section.

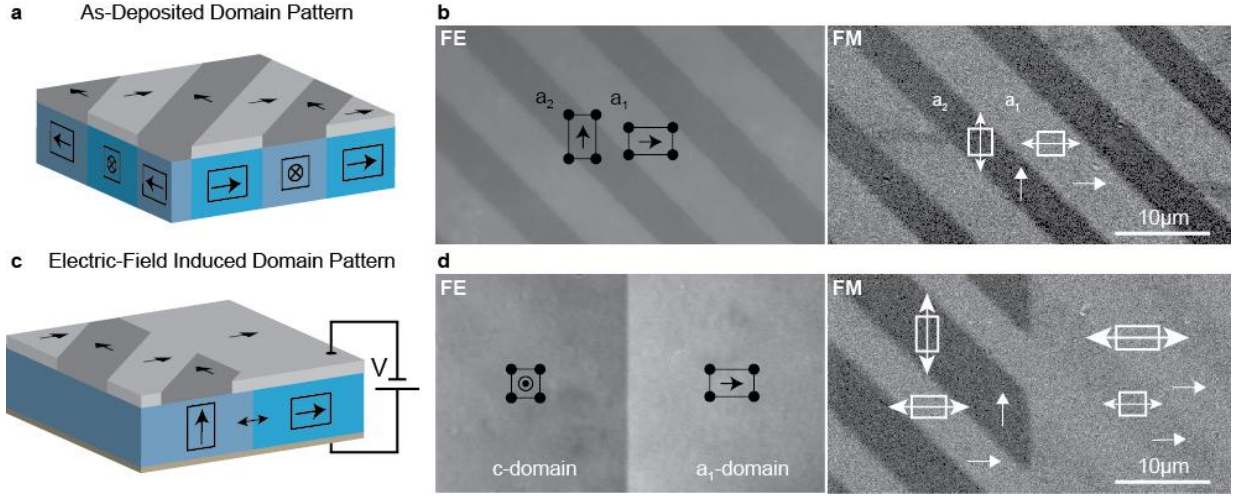

**Figure 1 Growth- and electric-field control of ferromagnetic domain patterns. a-d**, Schematic diagrams of the ferroelectric (FE) and ferromagnetic (FM) microstructure after CoFe film growth on $BaTiO_3$ (**a**) and the application of an out-of-plane electric field of 10 kV/cm (**c**) and corresponding polarization microscopy images (**b**,**d**). The polarization direction and lattice elongation of the $BaTiO_3$ substrate (black rectangles with arrows), the orientation of the strain-induced magnetic easy axis (white rectangles with double-headed arrows), and the magnetization direction in zero applied magnetic field (white arrows) are indicated.

Full imprinting of ferroelastic $BaTiO_3$ domains into CoFe during thin-film growth is demonstrated in Fig. 1b. We find exact correlations between both ferroic domain patterns across the entire sample area. Analysis of the birefringent contrast in FE and the magneto-optical Kerr contrast in FM indicates that the ferroelectric polarization and magnetization directions are collinear in zero magnetic field (black and white arrows). Precise reproduction of the ferroelastic domain structure in the CoFe film is explained by strain transfer at the heterostructure interface, which induces local modulations in the uniaxial magnetoelastic anisotropy via inverse magnetostriction. Next, we analyze the micromagnetic response to an out-of-plane electric field ($E = 10$ kV/cm) across the $BaTiO_3$ substrate. Figure 1d compares the ferroelectric and ferromagnetic domain pattern after reverting back to electrical remanence ($E = 0$ kV/cm). The images provide clear evidence of highly correlated changes in both ferroic domain structures. In the $BaTiO_3$ substrate, the electric field transforms the ferroelectric microstructure into alternating $a_1$ and $c$ domains as manifested by a 45° rotation of the ferroelectric domain boundaries. These ferroelastic modifications are accompanied by the transfer of local strains to the CoFe film, which, in turn, results in the erasure (on top of the $a_1$ domains) and conservation (on top of $c$ domains) of the original magnetic stripe pattern. The schematic diagrams in Fig. 1 illustrate the correlations between the two ferroic microstructures in the as-deposited and electric-field induced states.

To interpret the evolution of the ferromagnetic domain pattern, we need to first unravel the strain-mediated relationship between local magnetic anisotropies and the underlying ferroelectric domain structure. For this purpose, we utilized our polarization microscope to measure local magnetic hysteresis curves on single domain areas. Data as a function of in-plane magnetic field angle were used to determine the orientation of the uniaxial magnetic easy axis for each domain type. Moreover, the magnitude of the magnetoelastic anisotropy ($K_{me}$) was extracted by fitting the slope of hard-axis magnetization curves to $M_s/2K_{me}$, where $M_s$ is the saturation magnetization of CoFe ($1.7 \times 10^6$ A/m). Figure 2 summarizes the results for all ferromagnetic-ferroelectric domain combinations.

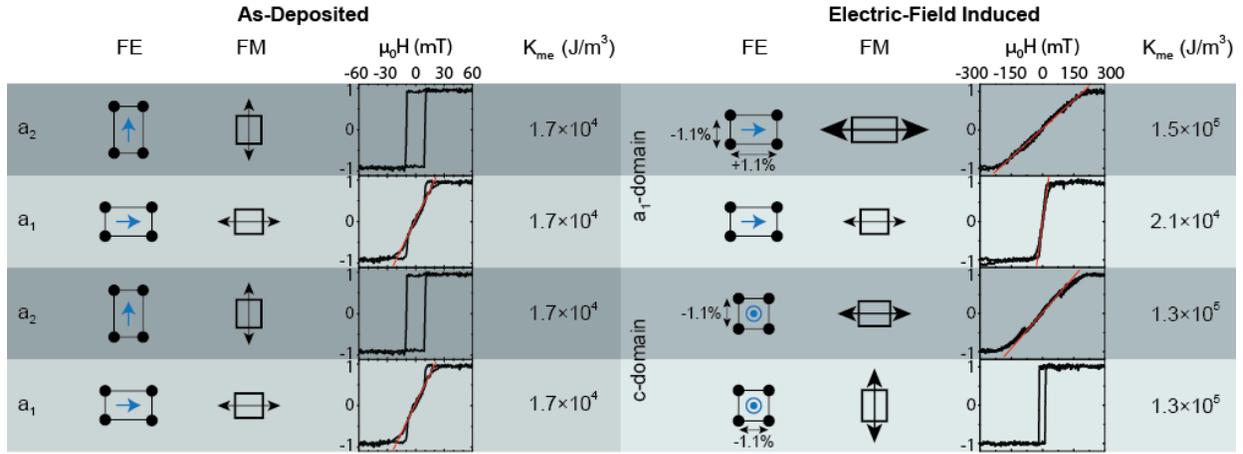

**Figure 2. Strain-mediated relationship between local magnetic anisotropies in the CoFe film and the underlying ferroelectric domain structure of the BaTiO$_3$ substrate.** The displayed hysteresis curves are measured on single magnetic domain areas with the in-plane magnetic field along the vertical axis of all images in this paper. The solid red lines are linear fits to the hard-axis magnetization curves.

After growth, the ferromagnetic microstructure consists of alternating $a_1$ and $a_2$ domains with horizontal and vertical magnetic easy axes and $K_{me} = 1.7 \times 10^4$ J/m$^3$. This anisotropy value can be put into perspective by comparing it to a theoretical estimation. For polycrystalline films, $K_{me} = 3\sigma\lambda/2$, where $\lambda$ is the magnetostriction constant and $\sigma$ is the stress, which is proportional to the strain $\varepsilon$ via Young's modulus $Y$. Using $\varepsilon = 1.1\%$ for full strain transfer from the BaTiO$_3$ substrate to the CoFe film and $Y = 2.5 \times 10^{11}$ J/m$^3$ [29] and $\lambda = 6.8 \times 10^{-5}$ [28] for CoFe, gives $K_{me, max} = 2.8 \times 10^5$ J/m$^3$. This calculation suggests that less than 10% of the BaTiO$_3$ lattice strain is transferred during CoFe film growth. Yet, our experiments clearly demonstrate that such limited strain effect is sufficient to fully imprint ferroelastic domains into magnetostrictive films when other anisotropy contributions are small.

Figure 1d shows that the magnetic microstructure consists of four different domain states after the application of an out-of-plane electric field. On top of the ferroelectric $a_1$ domains, new compressive and tensile strains in the original $a_2$ stripes of the CoFe film forced the uniaxial magnetic easy axis to rotate by 90°. As a result, the easy magnetization directions of neighboring magnetic stripes align parallel and, thus, the growth-induced domain pattern is electrically erased. The magnitude of the uniaxial magnetic anisotropy in the original $a_2$ stripes is found to increase significantly to $K_{me} = 1.5 \times 10^5$ J/m$^3$. This value compares more favorably to the theoretical $K_{me, max}$ and it clearly indicates that strain transfer is more efficient when the CoFe film is clamped to the BaTiO$_3$ substrate (as compared to the initial stages of film growth). For the magnetic $a_1$ stripes, the underlying ferroelectric lattice symmetry did not change, hence, $K_{me}$ remains small ($2.1 \times 10^4$ J/m$^3$).

Similar arguments hold for the development of magnetic anisotropies on top of the out-of-plane ferroelectric $c$ domains. In this case, the ferroelectric domain exhibits cubic symmetry in the substrate plane. Consequently, the CoFe film is compressed along the same axis along which it obtained a small tensile strain during deposition. As the electric-field controlled lattice compression in the CoFe film exceeds the growth-induced strain, the uniaxial magnetic easy axis rotates by 90° in both stripe domains and the strength of the magnetic anisotropy is found to increase to $K_{me} = 1.3 \times 10^5$ J/m$^3$. The original magnetic $a_1$ and $a_2$ domains thus transform to $a_2$ and $a_1$ stripes, respectively, but the pattern remains the same. Electric-field induced formation of ferroelectric $c$ domains thus results in the conservation and stabilization of the original magnetic domain structure.

We now demonstrate that strain-mediated ferroic domain correlations in CoFe-BaTiO$_3$ can be used to write and erase regular ferromagnetic domain patterns in small electric fields. An example is shown in Fig. 3, which depicts the evolution of the magnetic domain structure as a function of out-of-plane bias voltage. First, the ferroelectric polarization of the BaTiO$_3$ substrate is saturated by a bias voltage of 120 V (Fig. 3a). As previously discussed, this uniform $c$ domain produces a regular magnetic stripe pattern in the CoFe film. Reverting back to zero bias causes the ferroelectric microstructure to relax into alternating $a_1$ and $c$ domains (Fig. 3b). Consequently, the local strain state in the CoFe film and, thereby, the magnetoelastic anisotropies alter in accordance with Fig. 2. This results in the erasure of magnetic stripes on top of the ferroelectric $a_1$ domains. Subsequent application of a small bias voltage causes the ferroelectric $c$ domains to grow at the expense of $a_1$ domains by sideways motion of the ferroelastic $a_1 - c$ domain boundaries. This again modifies the local strain states in the CoFe film and the regular magnetic stripe pattern is electrically rewritten. The images of Fig. 3 indicate that electric writing and erasure of magnetic patterns based on strain-mediated correlations between local magnetic anisotropies and underlying ferroelastic domains is reversible. We note that this unprecedented electric-field control of ferromagnetic microstructures was obtained at room temperature without the application of a magnetic field.

Macroscopically, interlinking between ferromagnetic and ferroelectric order parameters is often quantified by the converse magnetoelectric coupling coefficient, $\alpha = \mu_0 \Delta M/\Delta E$. In our study, this coefficient is estimated from the electric-field induced change in the sample area that is covered by $a_1$ domains ($\Delta a_1$). At 0 V, the $a_1$ domains cover about 65% of the CoFe film and this reduces to 0% at 120 V ($E = 2.4$ kV/cm). By also taking into account that half of the magnetization rotates by 90° when $a_1$ domains are replaced by $c$ domains, we find $\alpha = 0.5\mu_0 M_s \Delta a_1/\Delta E = 3 \times 10^{-6}$ s/m, which is about one order of magnitude larger than previously reported results [19,24]. Our experiments thus clearly indicate that giant magnetoelectric coupling effects can be realized by careful optimization of the multiferroic microstructure.

Another intriguing observation in our study is electric-field control of magnetic domain wall motion. In this paragraph we analyze the main effects and the underlying driving mechanism. Figure 4a shows an enlarged image of the ferromagnetic domain pattern on top of a ferroelectric $c$ and $a_1$ domain. Two different types of magnetic domain walls are distinguished. The walls labeled as (1), appear on top of the ferroelectric $c$ domain and their location is determined by the position of $a_1 - a_2$ boundaries in the BaTiO$_3$ substrate during CoFe film growth. Thus, although this type of magnetic wall is not directly pinned onto an existing ferroelastic boundary, its physical properties (location, width, and pinning strength) are determined by local magnetic anisotropies that result from a previous ferroelastic domain configuration. This observation underpins the importance of ferroelastic domain history on magnetoelectric coupling effects. In contrast, the magnetic domain walls labeled as (2) in Fig. 4a are pinned on top of an existing ferroelastic $a_1 - c$ domain boundary. Sideways motion of such ferroelastic boundaries in an out-of-plane electric field results in the reappearance of type-1 walls at pre-defined locations, while type-2 walls are dragged along by their ferroelectric counterpart (Fig. 3). To analyze the properties of magnetic domain walls in more detail, we performed micromagnetic simulations using the local magnetic anisotropies of Fig. 2 as input parameters (Fig. 4b). From these simulations, the spin rotation within the domain wall and the pinning energy are extracted. The results indicate that type-1 and type-2 walls are strongly pinned onto previous and existing ferroelastic boundaries, respectively (Fig. 4c-d). The pinning energies provide robust wall localization and accurate electric-field control of magnetic domain wall motion. We note that despite these promising results, our observations are still early in our understanding of the details of ferromagnetic-ferroelectric domain wall pinning. Future experimental and theoretical studies are necessary to carefully examine the dynamics of such interlinked ferroic walls in an applied electric field and the ability to transfer these phenomena to well-defined nanostructures.

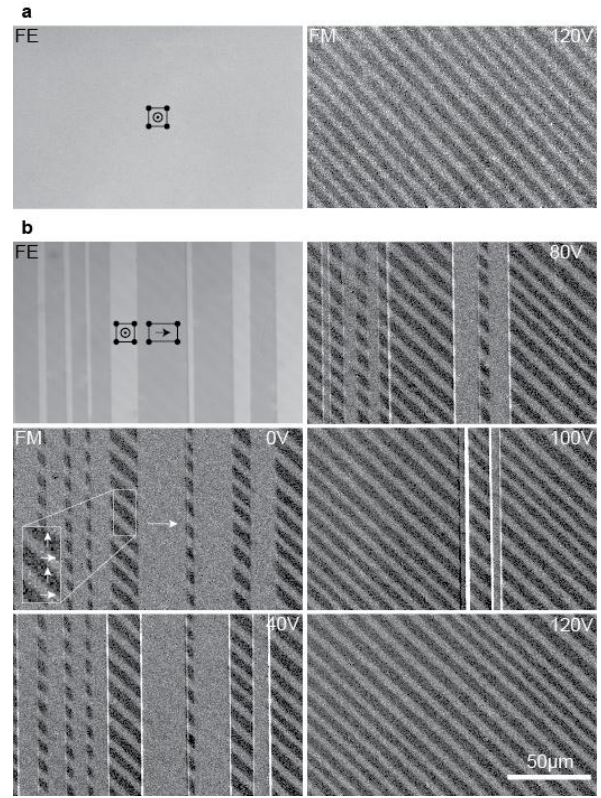

**Figure 3. Electric-field controlled writing and erasure of ferromagnetic domain patterns. a-b**, Polarization microscopy images of the ferroelectric (FE) and ferromagnetic (FM) domain structure during the application of an out-of-plane bias voltage. Correlations between the ferromagnetic microstructure and the underlying ferroelectric lattice are evident.

In summary, we have shown that strain transfer from ferroelastic domains in ferroelectric media to magnetostrictive films can be used to write ferromagnetic domain patterns and control magnetic domain wall motion by the application of an electric field at room temperature. The realization of these local magnetoelectric effects is an important step towards low-power spintronic devices.

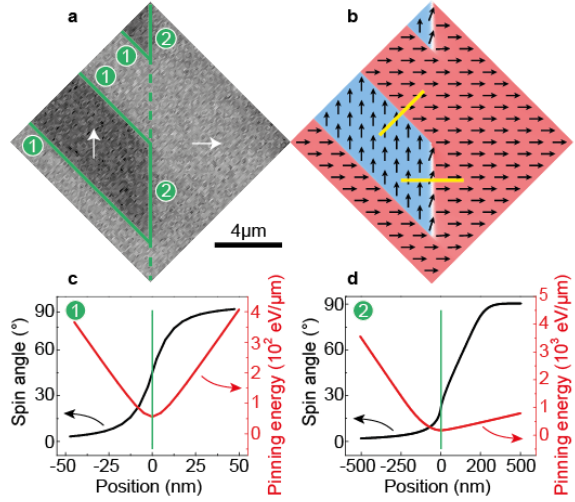

**Figure 4. Electric-field induced motion of magnetic domain walls. a**, Polarization microscopy image of the ferromagnetic domain structure on top of ferroelectric $c$ (left) and $a_1$ (right) domains. Two types of magnetic domain walls (numbered by (1) and (2)) and the underlying $a_1 - c$ domain boundary are indicated by solid and dashed lines, respectively. **b**, Micromagnetic simulation of the magnetic microstructure in **a**. **c-d**, Spin rotation within the magnetic domain walls and pinning energies indicating strong clamping of ferromagnetic walls onto ferroelastic boundaries (or magnetic anisotropy distributions that are the results of such boundaries). The pinning energy is calculated per μm domain wall length. The asymmetry in the wall profile and pinning energy in **d** is due to different magnetic anisotropy strengths on the left ($K_{me} = 1.3 \times 10^5$ J/m$^3$) and right ($K_{me} = 2.1 \times 10^4$ J/m$^3$) side of the ferroelastic $a_1 - c$ domain boundary.

## Methods

Thin CoFe films with an atomic percentage of 60% Co and 40% Fe were directly grown onto ferroelectric BaTiO$_3$ substrates by electron-beam evaporation at room temperature. Each film was covered by a 3 nm thick Au capping layer to prevent oxidation under atmospheric conditions. The thickness of the CoFe films was 15 nm and x-ray diffraction, atomic force microscopy, and transmission electron microscopy measurements confirmed the growth of smooth polycrystalline layers with a small root mean square roughness of <0.5 nm. The 0.5 mm thick BaTiO$_3$ substrates contained a regular $a_1 - a_2$ domain stripe pattern during CoFe film growth. The width of the ferroelastic stripes varied from 3 – 7 μm.

The ferromagnetic and ferroelectric domains were imaged independently using polarization microscopy techniques. The microscope consisted of polarizing optics, an adjustable diaphragm, a 100× objective, a CCD camera and an electromagnet for the application of in-plane magnetic fields. The ferroelectric BaTiO$_3$ domains were imaged using birefringent contrast with the polarizer and analyzer in cross-configuration. During these measurements the CoFe film was saturated by an in-plane magnetic field to eliminate magneto-optical contrast from overlying magnetic domains. The magnetic domain patterns were measured using the magneto-optical Kerr effect. In this case, the polarizing optics was set to a few degrees from extinction (to optimize the magneto-optical contrast) and a background subtraction method was used to zero-out contrast from structural defects and ferroelectric domains. Local magnetic hysteresis curves were constructed from the evolution of magnetic contrast in single domain areas during magnetization reversal. All ferromagnetic domain images in this paper (including those recorded as a function of out-of-plane bias voltage) were obtained at room temperature and zero applied magnetic field. Electric fields were generated by using the CoFe/Au film as top electrode and silver paint on the back of the BaTiO$_3$ substrates as bottom electrode. The electric bias voltage was ramped at 10 – 20 V/minute and a 20× objective was used to allow for wiring space during the poling experiment (Fig. 3).

The micromagnetic simulations (Fig. 4) were conducted using object oriented micromagnetic framework (OOMMF) software [30]. To closely mimic the experiments, the local magnetoeleastic anisotropies of Fig. 2 were used for the different magnetic domains. Other input parameters included a saturation magnetization of $M_s = 1.7 \times 10^6$ A/m and a uniform exchange constant of $K_{ex} = 2.1 \times 10^{-11}$ J/m. The domain wall pinning potentials were calculated from the orientation of magnetic spins along a line perpendicular to the walls using $U(x) = lt \int_{-\infty}^{\infty} K_{me}(x) \sin^2(\theta(x) - \alpha(x))\, dx$, where $K_{me}(x)$, $\theta(x)$, and $\alpha(x)$ are the strain-induced magnetoeleastic anisotropy, the angle of the spin moment, and the orientation of the uniaxial magnetic easy axis as a function of position, respectively. The parameters $l$ and $t$ indicate the length of the domain wall and the film thickness. In Figs. 4c and 4d, the pinning potential is calculated per μm domain wall length, i.e. $l$ = 1 μm.

## Acknowledgements

The work was supported by the academy of Finland under contract no. 127731. T.H.E.L. and K.J.A.F. also like to acknowledge support from the National Doctoral Program in Materials Physics and